\documentclass[
aps,
prb,
groupedaddress,
superscriptaddress,
floatfix,
twocolumn
]{revtex4-1}

\pdfoutput=1

\usepackage{graphicx}
\usepackage{amsmath,amssymb}
\usepackage{hyperref} 
\usepackage{bm} 
\usepackage{mathtools} 
\usepackage{float}
\usepackage{braket}

\usepackage{color}

\newif\iffigures
\figurestrue

\newcommand{\p}{\partial}

\begin{document}

\title{Mimicking graphene with polaritonic spin vortices}


\author{Dmitry R. Gulevich} \email{d.r.gulevich@metalab.ifmo.ru}
\affiliation{ITMO University, St. Petersburg 197101, Russia}

\author{Dmitry Yudin}
\affiliation{ITMO University, St. Petersburg 197101, Russia}

\date{\today}

\begin{abstract} 
Exploring the properties of strongly correlated systems through quantum simulation with photons, cold atoms or polaritons represents an active area of research. In fact, the latter permits to shed the light on the behavior of complex systems which are hardly to be addressed in the laboratory or tackled numerically. In this study we discuss an analogue of graphene formed by exciton-polariton spin vortices arranged into a hexagonal lattice. We show how the graphene-type dispersion at different energy scales arises for several types of exciton-polariton spin vortices. In contrast to previous studies of exciton-polaritons in artificial lattices, the use of exciton-polariton spin vortex modes offers a more rich playground for quantum simulations. In particular, we demonstrate that the sign of the nearest neighbor coupling strength can be inverted.
\end{abstract}


\maketitle

\textit{Introduction.}
Investigation of low dimensional strongly correlated electron systems remains one of the most exciting field of modern condensed matter physics. Nevertheless, only a limited number of physically relevant problems can be accomplished in a closed analytical form. Otherwise, some meaningful information about the nature of the ground states and correlation functions can be gained by making use of quantum simulators -- systems which, while being different physical systems, are described by the same or similar equations on the mathematical level. Of particular importance and interest are quantum simulators based on exciton-polaritons, a quasiparticle emerging from the strong light-matter coupling that simultaneously inherits the properties from both excitons and photons. Exciton-polaritons can be confined in artificially fabricated lattices and thus used to model the properties of crystalline solids.

Exploring typical features of exciton-polaritons confined within two-dimensional lattices represents an active area of research. 
In fact, only in the past few years, properties of polaritonic honeycomb lattice~\cite{Jacqmin-PRL-2014, Kusudo-2013, Milicevic-2015,Nalitov-graphene,Nalitov-Z,Bleu-Hall,Bleu-valley}, benzene molecule~\cite{Sala-PRX-2015}, kagome~\cite{Gulevich-kagome, Gulevich-SciRep} and Lieb~\cite{Lieb} lattices were addressed theoretically and experimentally, emergence of nontrivial topological phases of exciton-polaritons was discussed~\cite{Karzig-PRX-2015, Bardyn-PRB-2015, Yi-PRB-2016, Vasco-APL-2011, Kibble-Zurek}.
The use of exciton-polaritons allow to study condensed-matter phenomena in a well controllable environment where parameters of the systems can be changed by varying the confining potential landscape, changing the effective mass of a polariton, frequency of excitations, separation between the lattice sites. In this Rapid Communication we show that the sign of the coupling strength $J$ can also be controlled by involving higher excited states of the $p$-band in the form of polaritonic spin vortices.

As was first shown by Abrikosov \cite{Abrikosov} when a magnetic field is present it penetrates the superconducting condensate forming a triangular lattice of vortices. Each vortex carries a magnetic flux and its core represents a non-superconducting phase. The formation of vortices also has been shown to take place in Bose-Einstein condensates (BECs) of ultracold quantum atoms \cite{Madison2000,Abo2001,Bloch2008} and exciton-polaritons \cite{Lagoudakis2008,Sanvitto2010}. The formalism developed for vortex matter in superconductors can be easily generalized to magnetic skyrmion systems \cite{Bogdanov1989}. Skyrmions are characterized by a quantized topological winding number and are known to condense in a lattice quite similar to that of Abrikosov vortex lattice in superconductors \cite{Muhlbauer2009,Yu2010,Han2010}.

The first experimental observation of vortices in exciton-polariton systems in the regime of stationary incoherent pumping has been reported in Ref. \cite{Lagoudakis2008}. Interestingly, in contrast to BEC of ultracold atoms, the formation of vortices of exciton-polaritons is associated with presence of disorder in any realistic semiconductor heterostructure and non-equilibrium nature of the condensate. The optimal geometry of a vortex lattice is purely determined by the confining potential landscape in which polariton liquid is trapped. One more possibility to excite exciton-polaritons vortices is using a light beam with properly adjusted angular momentum \cite{Sanvitto2010}. In the field of confining potential the exciton-polariton liquid is no longer uniform, but at $T=0$ can be described by the spinor analogue of the Gross-Pitaevskii equation for the wave function of a condensate. Solutions to this equation are known to describe a whole bunch of phenomena ranging from simple equilibrium configurations to highly inhomogeneous vortices, bright and dark solitons~\cite{Aftalion2006}.




\textit{Polaritonic spin vortices.}
Recently the use of tunable open-access microcavities (OAMs)~\cite{Duff-APL-2014, Duff-PRL-2015} 
led to the observation of spin textures of exciton-polaritons -- polaritonic spin vortices (PSVs). This became possible owing to a large TE-TM splitting observed in OAMs which is reported to exceed those in monolithic cavities by a factor of three~\cite{Duff-PRL-2015}. The TE-TM splitting of exciton-polariton systems arises mainly due to the photonic component. The polarization splitting of exciton-polaritons results from the polarization-dependent reflection off dielectric mirrors in microcavity resonators~\cite{Panzarini-PRB-1999} which
manifests itself in the difference between the dispersion relations for polaritons polarized longitudinally and transversely to the direction of propagation (similar, yet a more delicate, effect also arises due to the excitonic component~\cite{Maialle-PRB-1993}). In the effective mass approximation the TE-TM splitting can be described by two masses $m_{\rm TM}$ and $m_{\rm TE}$ for polaritons polarized longitudinally (TM-mode) or transversely (TE-mode) with respect to 
the in-plane wave vector and 
in direct correspondence with the TE/TM resonator modes.
The TE-TM splitting in polariton systems leads to the emergence of the effective spin-orbit interaction~\cite{Kavokin-PRL-2005, Leyder-NP-2007, Sala-PRX-2015} which has been recently exploited in various settings related to topological properties of exciton-polaritons~\cite{Bardyn-PRB-2015, Nalitov-Z, Gulevich-kagome, Gulevich-spin-Meissner}.


Exciton-polaritons confined to a patterned structure with the potential energy landscape $V(x,y)$, are described by a system of coupled Gross-Pitaevskii equations~\cite{Flayac-PRB-2010,Solano-PRB-2014,Duff-PRL-2015}, 
\begin{equation}
\begin{dcases}
i\frac{\partial\psi_{+}}{\partial t} = -\Delta \psi_{+} + V(x,y)\,\psi_{+}
+ \beta \left(\partial_x - i \partial_y\right)^2 \psi_{-},
\\
i\frac{\partial\psi_{-}}{\partial t} = -\Delta \psi_{-} + V(x,y)\,\psi_{-}
+ \beta \left(\partial_x + i \partial_y\right)^2 \psi_{+},
\end{dcases}
\label{model}
\end{equation}
where
$\psi_\pm= (\psi_x \mp i \psi_y)/\sqrt{2}$ are two circular polarization components forming a spinor  $\psi = \psi_+ \hat{e}_+ + \psi_- \hat{e}_-$, with the basis vectors given by $\hat{e}_\pm = \left(\hat{e}_x \pm i \hat{e}_y\right)/\sqrt{2}$, $\beta$ is a parameter that corresponds to the TE-TM splitting~\cite{footnote}.
Here and in what follows we work with normalized units by introducing a characteristic length $L$ of the potential profile and related to it unit energy $E_0=\hbar^2/(2 m^*L^2)$,
on condition that $m^*=2m_{\rm TE}m_{\rm TM}/(m_{\rm TE}+m_{\rm TM})$ is the effective mass of a polariton. In the normalized units the TE-TM splitting parameter $\beta=(m_{\rm TE}^{-1}-m_{\rm TM}^{-1})m^*/2$.
Note that the sign of $\beta$ can be both positive and negative depending on the offset of the frequency from the center of the stop band of the distributed Bragg reflector~\cite{Panzarini-PRB-1999}. 

\begin{figure}
\begin{center}
\includegraphics[width=3.5in]{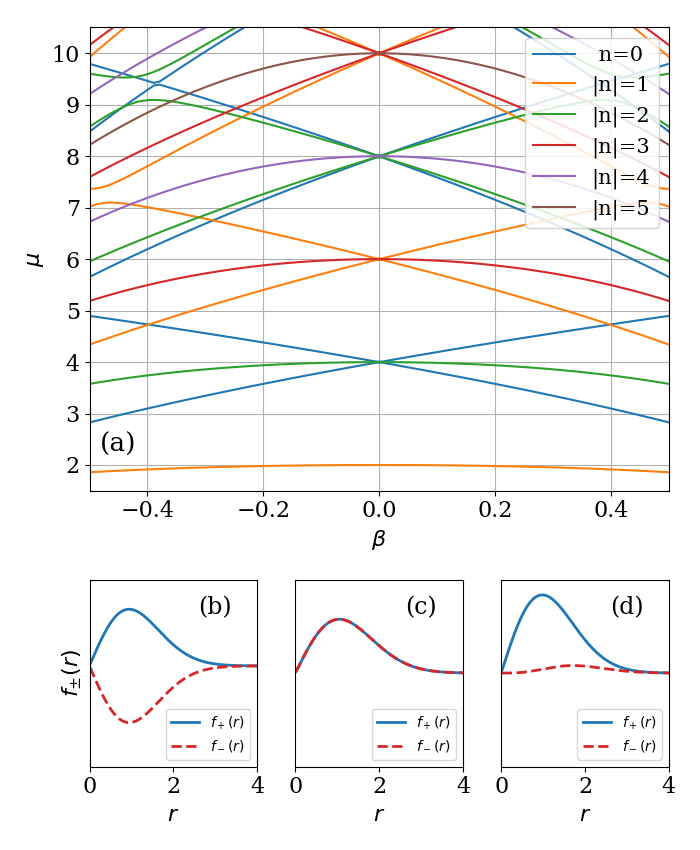}
\caption{\label{fig:spectrum} 
(a) Energy spectrum of polaritonic spin vortices in the harmonic potential given by the eigenvalue problem~\eqref{EVP}. Profiles of lowest energy spin vortices, depicted for TE-TM splitting $\beta=-0.2$, are shown in panels (b), (c), and (d). Panels (b) and (c) correspond to the two lowest energy states with $n=0$, $\mu=3.58$ and $\mu=4.38$, and (c) to the lowest energy state with $n=2$, $\mu=3.93$.
}
\end{center}
\end{figure}

In a typical experimental setup~\cite{Duff-PRL-2015} used to observe PSVs, the top concave mirror induces a strong and almost harmonic lateral confinement of polaritons.
Thus, mathematically, PSVs represent stationary solutions of the Eq.~\eqref{model} in the presence of the potential $V(x,y)=x^2+y^2$. 
Substituting the anzatz
\begin{equation}
\psi_\pm(r,\varphi) = e^{-i\mu t+i(n\mp 1)\varphi}f_{\pm}(r),
\label{anzatz}
\end{equation}
where 
$r$ is the radial distance,
to Eq.~\eqref{model} (compare~\eqref{anzatz} to solutions in a quasi-1D ring~\cite{Gulevich-spin-Meissner}) yields the equation on the envelopes $f_{\pm}(r)$,
\begin{equation}
\begin{split}
\mu f_\pm=\left[-\partial_r^2 - \frac{1}{r}\partial_r+\frac{(n\mp 1)^2}{r^2} + V(r)\right]f_\pm \\
\,+\, \beta \left[\p_r^2 + \frac{(1\pm 2n)}{r}\p_r + \frac{n^2 - 1}{r^2}\right]f_\mp.
\end{split}
\label{EVP}
\end{equation}
With requirements $f_\pm(r)$ behaves regularly at $r=0$ and is zero at infinity, $f(\infty)\to 0$, we obtain the 
spectrum by solving numerically the eigenvalue problem~\eqref{EVP}.
As seen from Fig.~\ref{fig:spectrum}a, if we take $\beta=0$ the spectrum coincides with that of Laguerre-Gauss modes. In the presence of a nonzero $\beta$, the degeneracy is lifted and the spectrum is split into individual modes. The spectrum is symmetric with respect to changing sign of $\beta$, although, the spin texture is not: two different types of spin textures are possible at the same energy when $\beta$ changes its sign.
Amplitudes $f_{\pm}(r)$ of the first few lowest energy PSVs are shown in Fig.~\ref{fig:spectrum}b,c,d at a chosen value of TE-TM splitting $\beta=-0.2$. It is clearly visible that
$|f_+(r)| = |f_-(r)|$ for the two $n=0$ vortices, which are, therefore, linearly polarized.
Introducing a linear polarization basis defined by $\psi_x=(\psi_+ + \psi_-)/\sqrt{2}$, $\psi_y=i(\psi_+ - \psi_-)/\sqrt{2}$, 
one can distinctly notice that the vortex in Fig.~\ref{fig:spectrum}(b) is polarized along the azimuthal direction, and
the vortex in Fig.~\ref{fig:spectrum}(c) is polarized along the radial direction (cf. Ref.~\cite{Duff-PRL-2015}).

\textit{Polaritonic spin vortex graphene.}
While the technology of fabricating artificial lattice of solid state microcavity seems to be well developed and actively applied to the creation of exciton-polaritons analogues of condensed matter systems~\cite{Schneider-2017}, 
coupling of multiple OAMs into an array is yet to be achieved. Dependence of the coupling strength of two OAMs on its separation has recently been studied in Ref.~\cite{Duff-APL-2015}.

A system of coupled OAMs arranged into an artificial lattice can be described by the system~\eqref{model} with the potential landscape
\begin{equation}
V(x,y) = 
V_0\, \min\limits_{i \in \Omega}\left[ 1, \frac{(x-x_i)^2+(y-y_i)^2}{R^2} \right],
\label{Vmodel}
\end{equation}
where $R$ is the radius of a single OAM and the minimum is taken over the set $\Omega$ of individual OAMs. 
To investigate the emergent behaviour of PSVs in lattices we focus on a honeycomb lattice. 
In our numerically calculation of the band structure we use the model~\eqref{Vmodel} with the cavity diameter matching the cavity separation (i.e., the touching cavities, which is an intermediate regime of coupling in the experimental study~\cite{Duff-APL-2015}), and parameters $m^*=2\times 10^{-5}\,m_e$, $\beta=-0.2$, the cavity diameter 4 $\rm\mu m$ and the potential strength $V_0=15\rm\;meV$ which are realistic for polaritonic systems.
A typical density distribution in a Bloch vector formed by the lowest energy PSV (the one presented in Fig.~\ref{fig:spectrum}b), is shown in Fig.~\ref{fig:bands}a. The complete $p$-band of the calculated band structure is shown in Fig.~\ref{fig:bands}. Therein, graphene-like dispersions are seen, arising due to each of the three types of PSVs. A nearly perfect graphene dispersion with well visible Dirac cones, at the top and bottom of the $p$-band in Fig.~\ref{fig:bands}, are produced by the lower (in energy) and the upper $n=0$ vortices (Fig.~\ref{fig:spectrum}b,c). The intermediate energies in Fig.~\ref{fig:bands} are occupied by the bands of the two $n=\pm 2$ vortices which are a doubly degenerate state in an isolated harmonic well. The finite intercavity coupling lifts the degeneracy and makes the band diagram more intricate than that of the $n=0$ vortices.

In contrast to previous studies of exciton-polaritons in artificial lattices, the use of exciton-polariton spin vortex modes offers a novel platform for emulating solid state systems, that is, the inverted sign of the nearest neighbor coupling strength. Indeed, in the conventional implementation of artificial lattices with exciton-polaritons, the coupling strength has been shown to be controlled by changing the landscape of the potential and separation between individual microcavity pillars or OAMs. However, the sign of the coupling parameter $J$ in the tight-binding Hamiltonian
$
\hat{H}=
- J \sum_{\langle ij\rangle} \left(\hat{a}_{i}^\dagger \hat{a}_{j}  + h.c.\right)
$
remains always positive for the $s$-band modes involved (it favors the neighbouring sites to match their wave function phases).
Using PSVs as a building block of artificial lattice systems, it is possible to achieve the opposite limit of negative  coupling strength, $J<0$. Therefore, PSVs provide further control over polaritonic systems and their application in modeling condensed matter phenomena.

\setlength{\unitlength}{0.1in}
\begin{figure}
\begin{center}
		$
		\begin{array}{c}				
		\begin{picture}(34,25)
		\put(2,0){\includegraphics[width=3.3in]{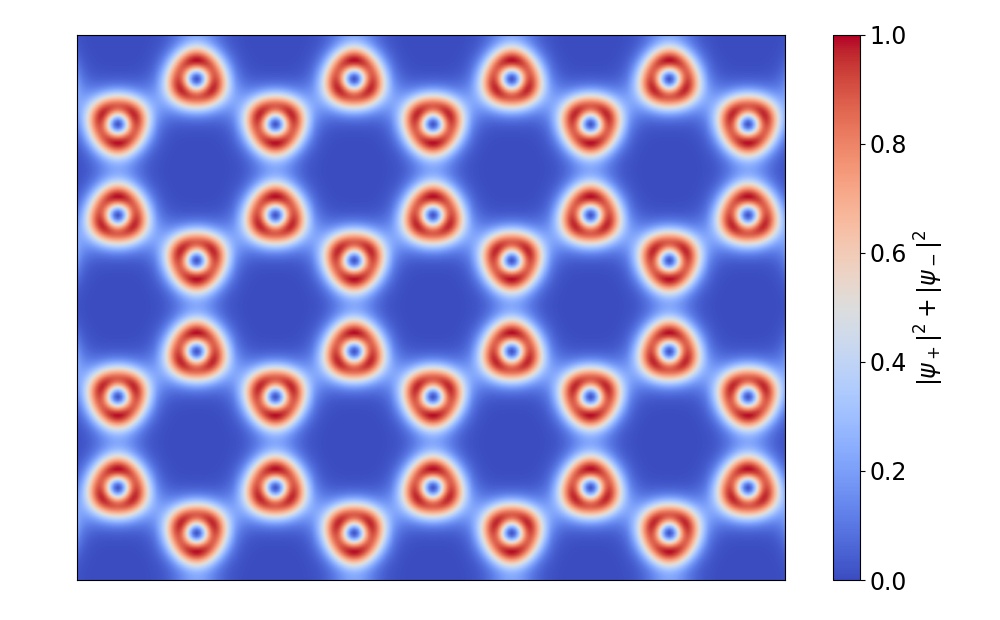}}
		\put(15,20){(a)}
		\end{picture}	
		\\
		\begin{picture}(34,25)
		\put(-1,0){\includegraphics[width=3.2in]{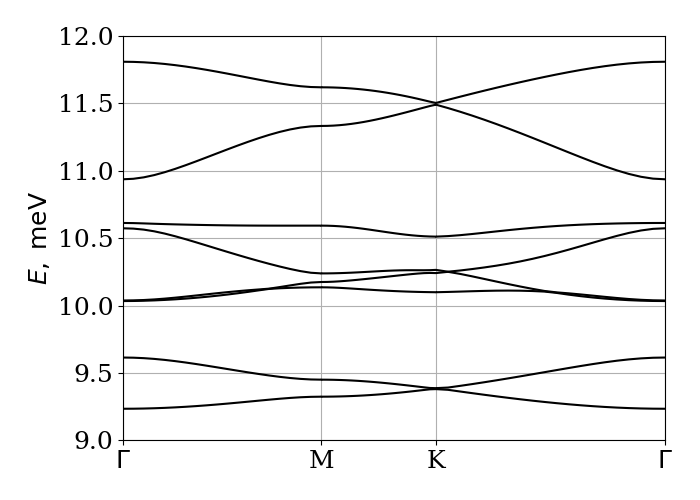}}
		\put(15,22){(b)}
		\end{picture}					
		\end{array}
		$	
\caption{\label{fig:bands} 
(a)
Numerically calculated Bloch state of polaritonic spin vortex graphene at $k_x = k_y = 0$. The color scale represents numerically calculated spatial distribution of exciton polariton density (arbitrary units) at $\beta=-0.2$.
(b)
Graphene-like dispersion formed by the lowest polaritonic spin vortices. Panels (a) and (b) are results of our numerical calculations with a 2D model~\eqref{model} and potential~\eqref{Vmodel} in the form of honeycomb lattice. Parameters $m^*=2\times 10^{-5}\,m_e$, $\beta=-0.2$, cavity diameter 4 $\rm\mu m$ and potential strength $V_0=15\rm\;meV$ were used in the calculation.
}
\end{center}
\end{figure}

\textit{Other types of polaritonic spin vortex lattices.}
Apart from the honeycomb lattice, it is instructive to look at the novel features brought by the use of PSV as fundamental modes of artificial lattices, such as, kagome lattice and a square lattice. 

Kagome lattice~\cite{kagome-PhysToday} is characterized by a high degree of frustration and three nonequivalent sites per a unit cell, has been attracting large interest in relation to topological properties of magnetic systems~\cite{Pereiro-2014, Chisnell-PRL-2015} and, very recently, exciton-polaritons~\cite{Masumoto-NJP-2012, Gulevich-kagome, Gulevich-SciRep}. In the tight-binding approximation with nearest neighbor coupling, the highest energy band of a polaritonic kagome lattice is completely flat. Arranging PSVs into a kagome lattice, allows to obtain the negative sign of the nearest neighbor coupling which results in the flat band being the lowest energy state of the whole $p$-band. 

With periodically arranged PSVs it is possible to deliver an analogy with magnetic systems. At $\beta<0$, the distribution of the electric field in a lowest energy PSV is  strictly azimuthal (it is perpendicular to the radial direction and oscillates between the clockwise and anticlockwise direction with polariton frequency).
According to the Maxwell equations, such configuration of electric field is associated with a magnetic field at the center of the vortex which is (i.e. transverse to the microcavity plane and oscillating between the up or down directions. 
Because two neighboring PSVs favor opposite phases, the oscillating magnetic field vectors in the neighbouring sites prefers an opposite orientation, that is, the antiferromengetic ordering. Indeed, if PSVs are arranged into a square lattice with the lattice spacing $a$, the energy minimum of the PSV configuration is achieved at the corners of the first Brillouin zone $k_x=k_y=\pm \pi/a$, in direct analogy with the antiferromagnetic arrangement of real magnetic moments.

\textit{Conclusions.}
We have demonstrated an analogue of graphene based on the PSVs. A well recognizable graphene dispersion arises for the several types of PSVs at different energy scales. Due to the non-equilibrium nature of exciton-polaritons, either of the dispersions can be observed by tuning the excitation frequency of the polariton system. With realistic parameters $m^*=2\times 10^{-5}\,m_e$, $\beta=-0.2$, the cavity diameter 4 $\rm\mu m$ and the potential strength $V_0=15\rm\;meV$ we obtained the two graphene dispersion band separation exceeding $500\rm\;\mu eV$ at the $\Gamma$ point which can in principle be observed experimentally.

We have also shown that the sign of the coupling strength~$J$ in polaritonic systems can be effectively controlled by employing PSVs.
In case of a kagome lattice the negative coupling strength $J$ results in the usual kagome's band structure to appear upside down, that is, the flat band being the lowest energy of the whole $p$-band. This fact is highly favorable for observation of exciton-polariton condensation and, in principle, may be exploited for engineering of polaritonic analogue of the quantum Hall effect.


\textit{Acknowledgments.}
We acknowledge the support from the Russian Science Foundation under the Project 17-12-01359.

\end{document}